\documentclass[prd,superscriptaddress,nofootinbib,showpacs,preprint]{revtex4}

\usepackage{graphicx}
\usepackage[usenames,dvipsnames]{color}
\usepackage{amsmath,amssymb,bbold}
\usepackage{hyperref}

\begin{document}

\title{Possibility of DCC formation in pp collisions at LHC 
energy via reaction-diffusion equation}

\author{Partha Bagchi}
\email{partha@iopb.res.in}
\author{Arpan Das}
\email{arpan@iopb.res.in}
\author{Srikumar Sengupta}
\email{srikumar@iopb.res.in}
\author{Ajit M. Srivastava}
\email{ajit@iopb.res.in}
\affiliation{Institute of Physics, Bhubaneswar, Odisha, India 751005}

\begin{abstract}
 There are indications of formation of a thermalized medium in high
 multiplicity pp collisions at LHC energy. It is possible that such a medium
 may reach high enough energy density/temperature so that a transient
 stage of quark-gluon plasma, where chiral symmetry is restored, may be 
 achieved. Due to rapid 3-dimensional
 expansion, the system will quickly cool undergoing spontaneous chiral
 symmetry breaking transition. We study the dynamics 
 of chiral field, after the symmetry breaking transition, for
 such an event using reaction-diffusion equation approach which we have
 recently applied for studying QCD transitions in relativistic heavy-ion 
 collisions. We show that the interior of such a rapidly expanding
 system is likely to lead to the formation of a single large domain of
 disoriented chiral condensate (DCC) which has been a subject of intensive
 search in earlier experiments. We argue that large multiplicity pp
 collisions naturally give rise to required boundary conditions for the 
 existence of slowly propagating front solutions of reaction-diffusion 
 equation with resulting dynamics of
 chiral field leading to the formation of a large DCC domain. 

\end{abstract}
\maketitle

\section{INTRODUCTION}

 Some years ago there was  a lot of interest in exploring the very
interesting possibility that extended regions, where the chiral field
is misaligned from the true vacuum, may form in large multiplicity 
hadronic collisions or in heavy-ion collisions \cite{anslm,bkt,blz,rw}. 
Such a region was called  a  disoriented chiral condensate (DCC). 
A large DCC domain would lead to spectacular signatures
such as coherent emission of pions which can be detected \cite{bkt} as
anomalous fluctuations in the ratio {\it R} of neutral pions to all
pions. Original motivation for DCC came from Centauro 
events in cosmic ray experiments \cite{cntr,bkt}. However, even after 
intensive experimental search for DCC, no clear signals were found for its
formation. Though it was generally agreed that in a heavy-ion collisions,
chiral symmetry breaking transition will necessarily lead to
formation of many DCC domains, expected size of such DCC domains was
too small, and their numbers too large in any given event, that 
standard DCC signals were washed out. Indeed, from this perspective,
heavy-ion collisions were not ideally suited for the  detection of DCC. 
With a large volume system undergoing chiral symmetry breaking transition,
multiple DCC domains necessarily result, and a clean signal of
coherent pion emissions becomes very unlikely. In comparison,
a pp collision, with a small volume system, could, in principle, 
lead to a single DCC domain. 

  We revisit the issue of formation of DCC, this time in the 
context of (very) large multiplicity pp collisions at LHC energies. 
Some of the earliest suggestions for DCC formation were actually 
made in the context of high multiplicity hadronic collisions. 
One would expect that a pp collisions, with a small volume system, 
could lead to a single DCC domain with a relatively cleaner signals
of coherent pion emission. However, at previously attained energies,
it was never clear whether the necessary condition for DCC formation, 
namely intermediate stage of chiral symmetry restoration, was ever 
achieved. Further, even if chiral symmetry was restored,
the resulting DCC domains would have been too small of order few
fm$^3$, in view of rapid roll down of the chiral field to the true 
vacuum. This will lead to only few pions from which a clear signal, 
say of neutral to charge particle ratio, would be hard to detect. 

The conditions of chiral symmetry restoration seem much more 
favorable for the very high multiplicity
pp collisions at LHC energy. Indeed there are strong indications that several 
signals, such as flow,  formation of ridge, etc. which have been
attributed to a thermalized medium undergoing hydrodynamic expansion in
heavy-ion collisions, may be present in such high energy pp 
collisions \cite{pplhc}. It is entirely possible that the energy 
density/temperature of such a medium may cross the chiral transition 
temperature. This will take care of the requirement of intermediate stage 
of chiral symmetry restoration for DCC formation. We show in this paper that
the problem of rapid roll down of the chiral field to true vacuum
is alleviated due to rapid three dimensional expansion of the system
which makes reaction-diffusion equation applicable for governing
the dynamics of chiral field for this system (with appropriate boundary
conditions which, as we will show, naturally arise in these events).
The expanding system leads to a DCC domain which stretches and becomes 
larger due to expansion, without the chiral field significantly
rolling down (due to specific properties of solutions of reaction-diffusion
equation). Eventually one gets a large DCC domain whose subsequent
decay should lead to coherent pion emission.

 We will not attempt to give any arguments in the favor of chiral
symmetry restoration in these high multiplicity pp events at LHC, and just
refer the reader to the literature where evidence for the possibility
of a thermalized medium in such collisions has been discussed \cite{pplhc}. 
We will only focus on the evolution of chiral field in such a system. 
Starting from a chirally symmetry phase (after some
very early stage of rapid thermalization of partons produced in a central
pp collision), rapid 3-dimensional expansion will quickly set in. This is 
due to the small size of the system resulting from pp collision, compared to
heavy-ion collisions where longitudinal expansion phase lasts for
significant time. Resulting rapid cooling of system will lead to
chiral symmetry breaking with the chiral field achieving some value
in the vacuum manifold. With explicit symmetry breaking term for the
chiral effective potential being small, any value in the vacuum manifold
will be (roughly) equally likely, leading to formation of a domain
where chiral field is likely to be initially misaligned from the true
vacuum. This will be a DCC domain. Standard estimates for such a domain
(from earlier investigations)  lead to typical size of coherence
length of order 1 fm. The field will also roll down to the true vacuum
rapidly in time of order few fm. It is very hard to detect such a DCC 
domain as this will lead to very small number of coherent pions.

  This is where the role of reaction-diffusion equation become
important. {\it Reaction-diffusion equations} \cite{reacdif,solns,hep},
are usually studied for biological systems, e.g. population 
genetics, and chemical systems. Interestingly, typical solution of such 
equations, with appropriate boundary conditions, consists of a traveling 
front with well defined profile, quite like the profile of the interface 
in a first order transition case \cite{reacdif,solns,hep}. This happens 
even when the underlying transition is a continuous transition or a 
crossover. In a previous work we have demonstrated that such {\it
propagating front} solutions, separating the two QCD phases, 
exist for chiral phase transition and
confinement-deconfinement (C-D) transition in QCD even when the underlying
transition is a cross-over or a continuous transition \cite{qcdrd}. 
We utilize the fact that  the only difference 
between the field equations in relativistic field theory case and the 
reaction-diffusion case is the absence of second order time derivative
in the latter case. Thus, correspondence between the two cases is
easily established in the presence of strong dissipation term leading to
a dominant first order time derivative term. Such a dissipative term
arises due to plasma expansion in the form of the Hubble term. 
Further, we had argued that the required boundary conditions for the 
existence of such a {\it traveling front} naturally arise in the context 
of relativistic heavy-ion collision experiments (RHICE). 

We extend that analysis \cite{qcdrd} to the case of high multiplicity pp 
collisions at LHC energy. As we are interested in the formation of DCC, we
focus here on the chiral transition. We argue that here also appropriate 
boundary conditions naturally arise which are suitable for the existence of 
propagating front solutions. One important difference between the analysis
in \cite{qcdrd} and the present case is that previously we considered
propagating front solutions separating chirally symmetric phase from the 
chiral symmetry broken phase. Here, in view of our focus on DCC formation, 
we consider the situation when the 
(approximate) chiral symmetry is spontaneously
broken after an early stage of chiral symmetry restoration. We then
consider the interior of the system to be such that chiral field is
{\it disoriented} there from the true vacuum, while it lies in
the true vacuum outside. This constitutes the initial profile of
the chiral field which gives the appropriate boundary conditions for
the existence of propagating front solutions for the reaction-diffusion
equation. We study evolution of this profile as the system undergoes 
rapid 3-dimensional expansion. The other requirement for the applicability
of reaction-diffusion approximation for this case is presence of strong 
dissipation. This is automatically satisfied due to dissipation term
(the Hubble term) arising from 3-dimensional spherical expansion
(as well as due to coupling of the chiral field with other field modes). 

  We will show that the propagating front solution {\it delays} the
roll down of the chiral field in the interior of the region towards the
true vacuum. At the same time rapid expansion stretches the interior
to a size of several fm radius before the field significantly
rolls down towards the true vacuum. Resulting system constitutes a
large, single, DCC domain which should lead to relatively clear
signal of coherent pion emission (e.g. in terms of the distribution of
neutral to charged pion ratio).

  We mention that in this paper we have ignored the effects of
thermal fluctuations. Such fluctuations are important and they will
lead to some variations in the chiral field within a domain.
However, in our model DCC formation results after chiral symmetry
breakdown when the system undergoes rapid 3-dimensional expansion,
hence rapid cooling. Thus, presumably, thermal fluctuations will 
remain under control. Main point is that large DCC domain here
results starting from a single small domain which stretches by rapid
expansion and the only role thermal fluctuations can play is to
fluctuate the field of this {\it single} DCC domain around the
average {\it disoriented} value. These considerations have to be
augmented with considerations of the quantum decay of the DCC domain
into pions which will put final limit on the growth of DCC domains
in our model.   

 The paper is organized in the following manner. In Sec. II, we provide
a brief review of DCC formation. Sec. III reviews basic physics
of reaction diffusion equations where we discuss that the dynamics of 
chiral order parameter for chiral symmetry breaking transition with 
dissipative dynamics is governed by one such  equation, specifically, 
the Newell-Whitehead equation \cite{qcdrd}. Sec.IV discusses the
basic physics of our model and Sec.V presents results
for the DCC formation. Conclusion are presented in section VI.

\section{Disoriented Chiral Condensate}

Formation of disoriented chiral condensates (DCC) in laboratory experiments 
was intensively investigated some time ago. DCC refers to
the formation of a chiral condensate in an extended domain, such that 
the direction of the condensate is misaligned from the true vacuum direction.
It is expected that as the chiral field relaxes to the true vacuum in such 
a domain, it will lead to coherent emission of pions. 
A motivation for the formation of such domains came from Centauro  events 
in cosmic ray collisions \cite{cntr}. It was suggested in 
ref.\onlinecite{bkt}  that the anomalous fluctuations in
neutral to charge pion ratio observed in the Centauro (and
anti-Centauro) events in cosmic ray collisions, could be due to the
formation of a large region of DCC. This was termed as the
{\it Baked Alaska} model in ref.\onlinecite{bkt}. Formation of DCC was 
extensively investigated in high multiplicity hadronic collisions as well
as in heavy-ion collisions \cite{anslm,bkt,blz,rw}.

A natural framework for the discussion of the formation of
DCC is the linear sigma model as this provides a simple way to 
model chiral symmetry restoration at high temperatures. Formation of 
DCC naturally happens as the temperature drops down through the critical
temperature, and the chiral field picks up random directions
in the vacuum manifold in different regions in the physical space.
We will work within the framework of linear sigma model with the
Lagrangian density given by,

\begin{equation}
L = {1 \over 2} \partial_\mu {\bf \Phi} \partial^\mu {\bf \Phi}
- V({\bf \Phi},T)
\end{equation}

where the finite temperature effective potential $V(\Phi,T)$ at 
one loop order is given by \cite{bynsk},

\begin{equation}
V = \frac{m_\sigma^2}{4} \left({T^2 \over T_c^2} - 1\right) 
|{\bf \Phi}|^2 + \lambda |{\bf \Phi}|^4 
- H \sigma
\end{equation}

\noindent Here the chiral field ${\bf \Phi}$ is an $O(4)$ vector
with components ${\bf \Phi} = ({\vec \pi},\sigma)$, and $T$ is the
temperature. Values of different parameters is taken as
$m_\sigma$ = 600 MeV, $\lambda \simeq 4.5$, $H$ = (120 MeV)$^3$
and $T_c \simeq 200$ MeV.

\begin{figure}[!htp]
\begin{center}
\includegraphics[width=0.6\textwidth]{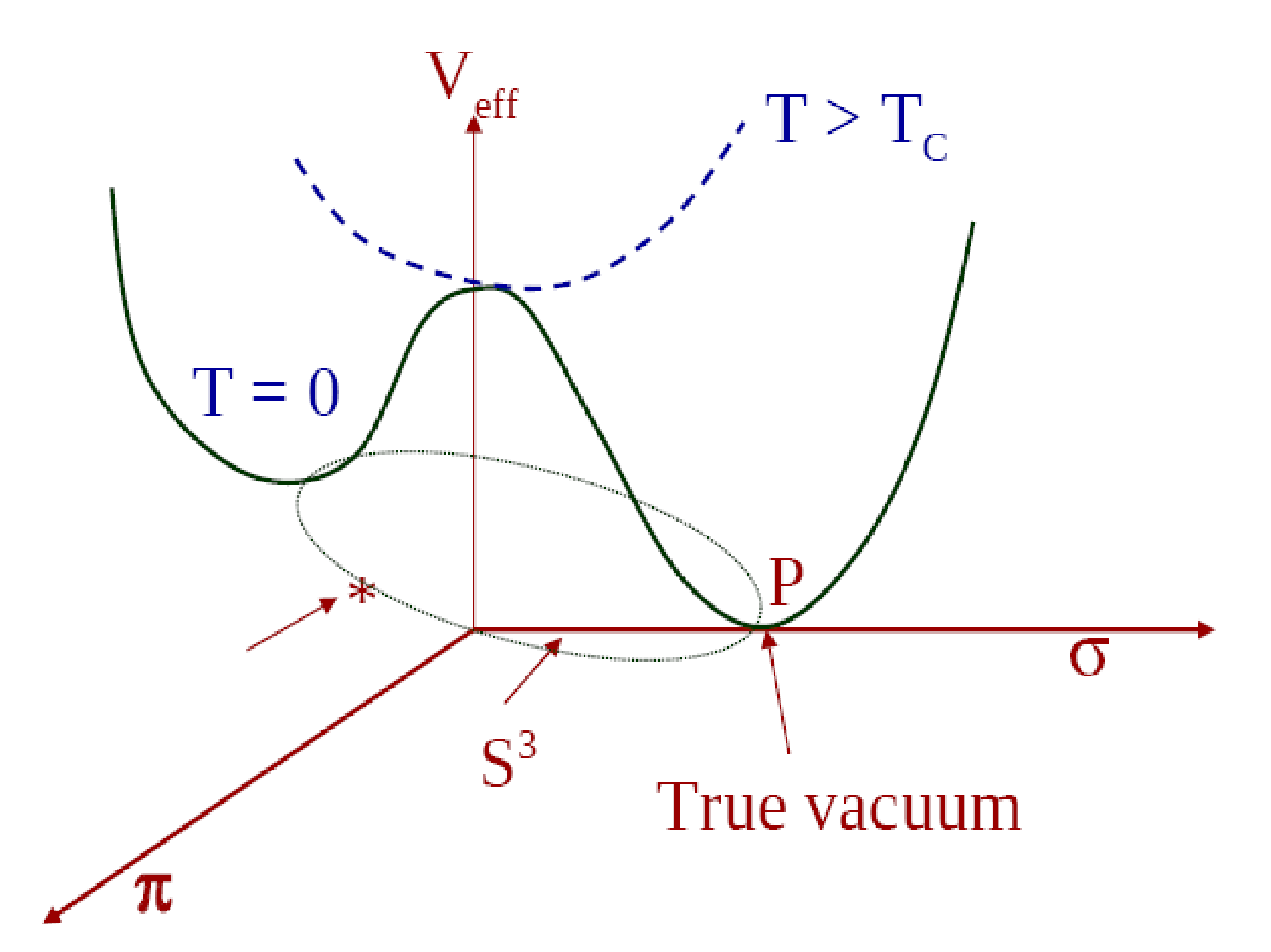}
\caption{Effective potential for the chiral field ${\bf \Phi}$.
$P$ denotes the true vacuum on the (approximately degenerate)
vacuum manifold while $*$ marks the value of the  chiral field
inside a DCC domain which is {\it disoriented} from the true 
vacuum direction.}
\label{fig1}
\end{center}
\end{figure}

In the chiral limit, spontaneous
breaking of chiral symmetry (for $T < T_c$) implies that the vacuum
corresponds to some specific point on the vacuum manifold $S^3$ 
with all points on $S^3$
being equally likely. This is not the situation in the presence of
explicit symmetry breaking, as there is a unique vacuum state
as shown in Fig.1. However, one may expect that this {\it preference}
for the true vacuum may be insignificant during very early stages
in a rapidly cooled system, due to small pion mass. Thus, as the temperature
drops below $T_c$, one expects that the chiral field will assume
some arbitrarily chosen value in the (approximately degenerate) 
vacuum manifold within a correlation size domain. If this value
differs from the true vacuum direction (as marked by $*$ in Fig.1)
then this domain will correspond to a DCC which will subsequently 
decay by emission of coherent pions as the chiral field rolls down 
to the true vacuum. This essentially summarizes the conventional picture 
of the formation of a DCC domain.

\section{REACTION-DIFFUSION  EQUATION FOR CHIRAL TRANSITION}

There is a wide veracity of reaction-diffusion equations, see, e.g.
ref.\cite{reacdif,solns}. Previously we discussed \cite{qcdrd} specific 
equations which can be identified with the  field equations for the chiral 
transition and the C-D transition in QCD in strong dissipation limit, leading
to slowly moving propagating front solutions.
We then showed that in different situations in relativistic heavy-ion
collisions, with realistic dissipation, propagating front solutions of 
these equations still persist, making the dynamics of the relevant 
transitions effectively like a first order phase transition. Here, we
will recall the case of chiral transition from ref.\cite{qcdrd} and
adopt it for the situation for the evolution of field inside a DCC domain. 

 As we mentioned in the introduction, we will consider the case of
high multiplicity pp collisions at LHC energy, assuming that
the resulting partonic system undergoes a rapid 3-dimensional
expansion. We  take the field equations for the chiral field 
(from Eq.(1)) to be,

$${\ddot \phi} - \bigtriangledown^2 \phi + \eta {\dot \phi} =
-4\lambda \phi^3 + m(T)^2\phi + H$$
\begin{equation}
m^2(T) = {m_\sigma^2 \over 2} (1 - {T^2 \over T_c^2})
\end{equation}

Here $\Phi$ is taken to be along $\sigma$ direction only which we
represent by $\phi$. This is
for the sake of establishing correspondence with the reaction-diffusion
equation. Later, when we consider the case of DCC, we will consider
other components of the chiral field $\Phi$ as well. In the above equation,
the time derivatives are w.r.t the proper time $\tau$.
The dissipation term $\eta$ is not a constant for expanding plasma. 
For the early stages in a heavy-ion collisions one normally takes 
Bjorken 1-D scaling solution case with $\eta = 1/\tau$, which eventually
turns into a 3-dimensional spherical expansion for which 
$\eta = 3/\tau$. For the present case of pp collision, due to small system 
size, one expects 3-dimensional expansion to be applicable from very
early stages (after time of order 1 fm). Hence, later on when we discuss 
the case of DCC, we will take $\eta = 3/\tau$.

Exact correspondence of the above equation with the reaction-diffusion 
equation was established in ref.\cite{qcdrd} by neglecting the explicit 
breaking of chiral symmetry (i.e. $H = 0$ in Eq.(1)) and by considering
the extreme dissipative case of large, constant, value of $\eta$  so
that one could neglect the ${\ddot \phi}$ term. With rescaling of the
variables as, $x \rightarrow m(T) x$, $\tau \rightarrow {m(T)^2 \over
\eta} \tau$, and $\phi \rightarrow 2{\sqrt{\lambda} \over m(T)} \phi$,
the resulting equation is found to be,

\begin{equation}
{\dot \phi} =  \bigtriangledown^2 \phi -\phi^3 + \phi
\end{equation}

This equation, in one dimension with
$\bigtriangledown^2 \phi = d^2\phi/dx^2$, is exactly
the same as the reaction-diffusion equation known as the Newell-Whitehead
equation \cite{reacdif,solns}. In that context, the term $d^2\phi/dx^2$ 
is identified as the diffusion term
while the other term on the right hand side of Eq.(4) is the so called {\it
reaction term} (representing reaction of members of biological species
for the biological systems). Non-trivial traveling front solutions for 
the Newell-Whitehead equation arise with suitable boundary conditions,
namely $\phi = 0$ and 1 at $x \rightarrow \pm \infty$.
The analytical solution with these boundary conditions has the form,

\begin{equation}
\phi(z) = [1+exp(z/\sqrt 2)]^{-1}
\end{equation}

where $z = x-v\tau$. $v$ is the 
velocity of the front \cite{solns} and has the value $v = 3/\sqrt 2$
for this solution.

 One can see from the general form of these reaction-diffusion equations,
that such traveling front solutions will exist when the underlying
potential allows for non-zero  order parameter in the vacuum state,
along with a local maximum of the potential \cite{reacdif,solns}. 
The corresponding values of the order parameter provide the required 
boundary conditions for the propagating front solution. In 
ref.\cite{qcdrd} we were interested in the dynamics of chiral symmetry
breaking transition, hence we considered the two boundary values of 
the chiral field to be the true vacuum value and the one corresponding
to the central maximum of the potential, respectively.
For the case with non-zero value of $H$ as in Eq.(3), the value of 
chiral field at one boundary was taken to be the (true) vacuum 
expectation value  $\phi = \xi$ while the other 
boundary field value corresponded to the shifted central 
maximum of the potential as $\phi = \phi_0$ (see, Fig.1). 
The propagating front solution in Eq.(5), suitably modified for 
these changed boundary conditions is \cite{qcdrd},

\begin{equation}
\phi(z) = - {(\xi - \phi_0) \over A_0} [1 + 
exp({m(T)(|z|-R_0) \over \sqrt{2}})]^{-1} + \xi
\end{equation}

where the normalization factor  $A_0 = [1 + exp({-m(T)R_0 \over 
\sqrt{2}}]^{-1}$. Here, we have restored the original, unscaled,
variable $z$. $|z|$ was used in order to have symmetric 
front on both sides of the plasma for the 1-d case with
$R_0$ representing the width of the central part of the plasma.
For the 3-dimensional case, $|z|$ is replaced by the radial
coordinate $r$. (Also, for the present case of pp collisions,
we will multiply $m(T)$ by 3 to represent a sharper variation of
$\phi$ initially. Note, this will be just a suitable choice
of initial profile and proper solution of propagating front will
result quickly when the initial profile is evolved by the field 
equations.) 

  In ref.\cite{qcdrd} we had calculated numerical solutions for the 
full Eq.(3), retaining
the ${\ddot \phi}$ term. Correspondence with the analytical solution
was achieved by considering a  large, constant, value of $\eta$ which
resulted in  propagating fronts of the same form as
discussed in literature for reaction-diffusion equations. Subsequently
we relaxed this assumption of constant $T$ and studied proper time dependence
of $T$ and $\eta$ for expanding QGP (still retaining the assumption of uniform 
temperature for studying front propagation as with spatially varying
$T$ the effective potential also has to vary spatially and correspondence
with reaction-diffusion equation becomes more complicated). 
We showed that the propagating front solution still exists with little
modifications.

\section{DCC formation via the reaction-diffusion equation}

In the present work we are considering the situation of the 
evolution of the {\it disoriented} chiral field after chiral symmetry 
breaking transition. Thus the two boundary conditions for the propagating
front solution have to be appropriately modified. The basic picture
of DCC formation in this case will be taken as follows. In a high 
multiplicity pp collision, we will assume that a thermalized medium is
created and that temperature/energy density of this medium reaches 
sufficiently large value so that the
(approximate) chiral symmetry is restored during very early stages.
Due to very small size of the initial system, it undergoes 3-dimensional
spherical expansion after a very short time of order 1-2 fm. This leads
to rapid cooling of the system and the chiral symmetry is spontaneously
broken. This is the starting point of our calculation, with initial
profile of the chiral field in the interior of the system assuming
some arbitrarily chosen value on the (approximately degenerate) 
vacuum manifold. We will consider the case of maximal disorientation
when the field in the center of the parton system takes value at the
saddle point {\it opposite} to the true vacuum on the vacuum manifold.
Outside the system the chiral field was always in the true vacuum
and we will assume that it continues to have close to the same value
in somewhat interior regions as well (where temperature could be taken
to develop similar value as the central temperature). This sets the two 
boundary conditions for the initial profile of the chiral field and we
study whether an initial profile with such boundary conditions can lead
to a propagating front solution. (We mention that for the sake of numerically
integration of the differential equation, we need to fix the boundary
conditions at $r = 0$ and for large $r$. However, the initial profile
taken has a plateau for small $r$ hence the field is allowed to
roll down freely in the region away from $r = 0$. Indeed, such a profile
with the same boundary conditions shows rapid roll down for the symmetry
restored potential where one does not expect any propagating
solution. Also, for one dimensional case, we consider the chiral field
to have symmetric profile about $x = 0$ and the boundary conditions are
only fixed for large $|x|$ with $x=0$ point free to evolve via
the differential equation. Exactly same results of propagating
front solution are still obtained on both sides of $x = 0$.)

The profile of the chiral field in between the two boundary values 
(as discussed above) is  taken to lie on the vacuum manifold and we 
choose this profile, for simplicity, to remain in 
the $\sigma -\pi_3$ plane. In such a DCC domain, the decay of the
field will lead to emission only of neutral pions. If we had chosen
the field to remain in any plane of ($\pi_1,\pi_2,\sigma$), it would
lead to emission only of charged pions. For a more general possibility,
appropriate distribution of neutral and charged pions will be obtained.

 Note that it is not immediately obvious that such boundary
conditions should lead to a propagating front solution. For 
reaction-diffusion equations, the corresponding boundary condition is
set for a local maximum of the potential, and not for a saddle point.
However, it appears that the importance of maximum of the potential is in 
delaying the roll down of the field from that point due to vanishing field
derivative. In that situation, a saddle point  will also 
satisfy this requirement and propagating front solution should result. 
As we will see, this intuition seems correct and we do find propagating 
solutions with this new type of boundary conditions. We have checked
that if the field at that boundary is taken even close  to the
saddle point (say, within 10-20 \%), slowly  propagating front still 
results and our results remain essentially unaffected. 

 For the present case of 3-dimensional expansion, with spherical
symmetry, will use the field equations in spherical polar coordinates,

\begin{equation}
{\ddot \Phi_i} - {d^2\Phi_i \over dr^2} - {2 \over r}{d\Phi_i \over dr}  
+ ({3 \over \tau} + \eta^\prime(T)) {\dot \Phi_i} = 
-4\lambda |\Phi|^2\Phi_i + m(T)^2\Phi_i + H \delta_{i4}
\end{equation}
 
where $\Phi_i$ denote components of the O(4) vector $\Phi$.   
For this 3-dimensional expansion case, the temperature is taken to
vary with proper time as,

\begin{equation}
T(\tau) = T_0 {\tau_0 \over \tau}
\end{equation}

The initial value of the temperature for field evolution is 
taken to be $T = T_0 = 150$ MeV, at proper time $\tau = \tau_0 = 2$ fm. 
This stage  corresponds to chiral symmetry broken phase. The system is
assumed to have reached a value larger than the critical temperature
at an earlier stage which allows for the chiral field to become
disoriented after the transition.  
Here, we have introduced a new dissipation parameter $\eta^\prime(T)$,
in addition to the Hubble damping coefficient $3/\tau$ \cite{lrgdcc,etap}. 
$\eta^{\prime}$ represents dissipation due to coupling to the heat bath,
or due to other field modes (which could be fields other than the chiral
field, or even high frequency modes of the chiral field itself).
The value of this dissipation parameter has been discussed in literature
(see, e.g. ref.  \cite{lrgdcc,etap} and references therein).  
We mention that inclusion of 
$\eta^\prime$ is not essential in our model of DCC formation as Hubble 
damping itself can be very large at sufficiently early times. However, from 
general considerations, one will always expect such an additional damping,
and it certainly helps for getting a slow moving propagating front
leading to a large DCC domain. We will first take $\eta^\prime \propto T^2$
with the initial value $\eta^\prime = $ 10 fm $^{-1}$ at 
$\tau = \tau_0$. Subsequently we will also consider the case  with
constant, time independent, $\eta^\prime = $ 20 fm$^{-1}$ 
and 40 fm$^{-1}$. We consider these
larger dissipation cases to allow for the possibility of the chiral field
coupling to other field modes, and to show that larger dissipation
can lead to much larger increase in DCC domain size in this model.

\section{Results}

 We now present results of field evolution via Eq.(7).
 The initial profile of the chiral field, at $\tau = \tau_0 = 2$ fm, 
is shown in Fig.2a. Solid curve shows the profile of the $\sigma$ field
which interpolates between the true vacuum value $\sigma = 75.18$ MeV
and the saddle point opposite to the true vacuum where $\sigma = 
-49.25$ MeV (with ${\vec \pi} = 0$ at both these boundaries). Interpolating
profile of $\sigma$ is taken in accordance with Eq.(6) (for the 3-dimensional
case with radial coordinate $r$ as discussed there), with 
$\phi_0$ and $\xi$ suitably replaced by the boundary conditions for
the present case. Further,
the chiral field is taken to lie everywhere on the (approximately
degenerate) vacuum manifold, hence the ${\vec \pi}$ field 
also varies in between the two boundary points, as shown by
the dashed curve for $\pi_3$ in Fig.2a . This is fundamentally
different from the case of chiral transition considered in our
previous work where pion field was taken to be zero all along the profile
of $\sigma$ which interpolated between the true vacuum and the central
maximum of the potential. We again mention that the choice of the
chiral field to lie entirely in the $\sigma - \pi_3$ plane is just
an example. Such a DCC will decay by emitting neutral pions. One
could take a more general variation, in which case an appropriate
distribution of neutral and charged pions will result.

We have taken the radius of the system to be about 2.5 fm assuming that 
the initial dense parton system in the pp collision would have
undergone some expansion by the time this stage is achieved at $\tau_0 = 2$ fm. 
This initial profile is evolved using Eq.(7). Note that Eq.(7) is written
in comoving coordinates. As the system is undergoing 3-dimensional scale
invariant expansion, the physical distances have to be obtained by 
multiplying with the appropriate scale factor. For this purpose we
have taken the velocity of the plasma at comoving distance $r$ to be 
proportional to $r$, with some maximum velocity at the boundary of
the region (which we take as a sample value to be 0.9).
Plots at subsequent stages are shown in 
Figs.2b-d with the x axis denoting the physical distance. This is where
we see the importance of the front solution of the reaction-diffusion 
equation. Normally one would have expected that the field from the 
saddle point will roll down towards the true vacuum in a time scale of
couple of fm within the whole system of size 2-3 fm. However,
the front solution delays this roll down dramatically. The field
retains its value close to the saddle point in a significant
region for a long duration of time (due to slow motion of the front). 

During this period, rapid expansion of the plasma
stretches the whole system, thereby stretching the region where
the chiral field is close to the saddle point, hence disoriented.
This leads to a DCC domain which is expanding and getting bigger
without the chiral field in the interior rolling down towards 
the true vacuum. This is shown in Fig.2b (for the $\sigma$ field)
and Fig.2c (for $\pi_3$). Note that stretching of a DCC domain
costs energy and this should be properly accounted for by calculating
back reaction of DCC stretching on the expanding plasma. However,
for ultra relativistic pp collisions the expanding parton system will
have very large kinetic energy, and the effects of back reaction 
of stretching of DCC domain will not be significant for the time scales
considered here.  Fig.2b shows the $\sigma$ field profile (dashed curve)
at $\tau = 4$ fm
clearly showing that the DCC domain (the region where the field
is significantly disoriented from the true vacuum) has almost
doubled in size. This means multiplication in number of coherent
pions by a factor of 8 (compared to the number expected from
the DCC domain of initial parton system size) when the DCC eventually 
decays. Fig.2d shows the situation at $\tau \simeq 7.2 $ fm when the
chiral field has significantly rolled down towards the true vacuum.
One can say that the decay of DCC domain has set in by
this stage. Eventually the DCC domain decays with chiral field rolling 
down to the true vacuum.   

\begin{figure}[!htp]
\begin{center}
\includegraphics[width=0.9\textwidth]{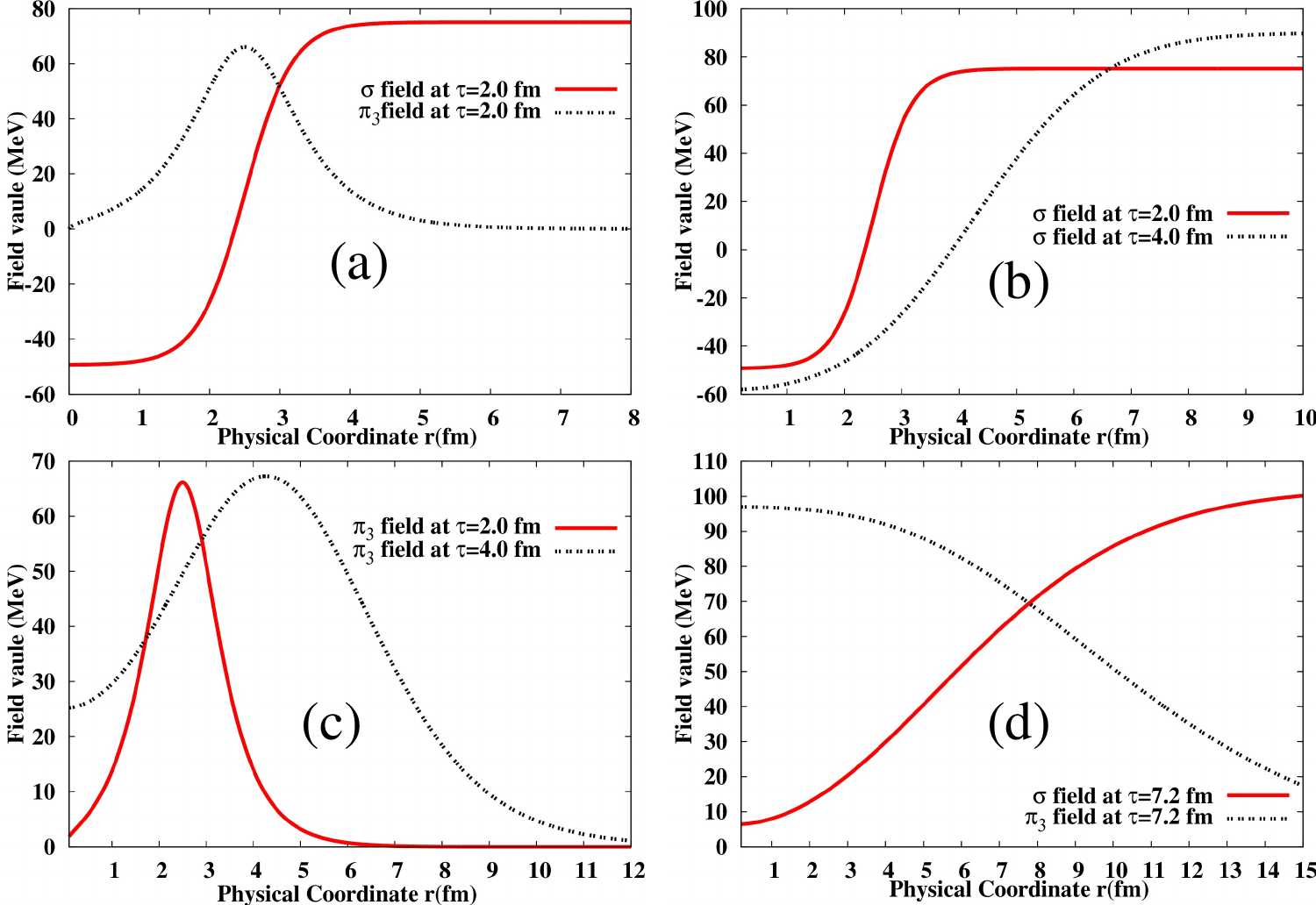}
\caption{(a) The initial profile of the chiral field. 
Solid (red) curve shows the profile of the $\sigma$ field
which interpolates between the true vacuum value $\sigma = 75.18$ MeV
and the saddle point opposite to the true vacuum where $\sigma = 
-49.25$ MeV. Corresponding variation of $\pi_3$, ensuring that
the field (approximately) lies on the vacuum manifold, is shown by the 
dashed (black) curve. (b) Dashed (black) curve shows the profile of 
$\sigma$ field after the system has undergone expansion up to $\tau = 4$ 
fm. For comparison, solid (red) curve shows the initial $\sigma$ 
profile. Stretching of the plasma leading to expansion of the DCC domain
is clearly seen. (c) Corresponding profile of
$\pi_3$ field at $\tau = 4$ fm is shown by the dashed (black curve), while
solid (red) curve shows the initial $\pi_3$ profile. (d) This shows the 
stage at $\tau \simeq 7.2$ fm when the decay of DCC domain has set in 
with chiral field significantly moving away from the initial disoriented 
value. Solid (red) and dashed (black) curves show, respectively,
$\sigma$ and $\pi_3$ field profiles.}
\label{fig2}
\end{center}
\end{figure}

 We now consider case of larger dissipation with constant $\eta^\prime$.
Figs.3a,b show the stages corresponding to the stages in Fig.2b,c,d for
the case with constant $\eta^\prime $ = 20 fm$^{-1}$. Figs.3c,d show 
similar stages for constant $\eta^\prime $ = 40 fm$^{-1}$. We note
significant increase in the stretching of DCC domains. In fact by
$\tau = $ 7.2 fm the fields have still not started significantly
deviating from the disoriented value. We do not show plots for
large $\tau$ because, as mentioned above, the decay of DCC domain 
by thermal fluctuations as well
as quantum effects will limit the growth of DCC domain.
It is not clear if  such large and (quasi) constant values
of $\eta^\prime$ can be realistic. However, its significant
effect on the formation of large DCC domains may be taken as a strong
motivation for finding arguments/situations where such strong 
dissipation may be justified/applicable. 

\begin{figure}[!htp]
\begin{center}
\includegraphics[width=0.9\textwidth]{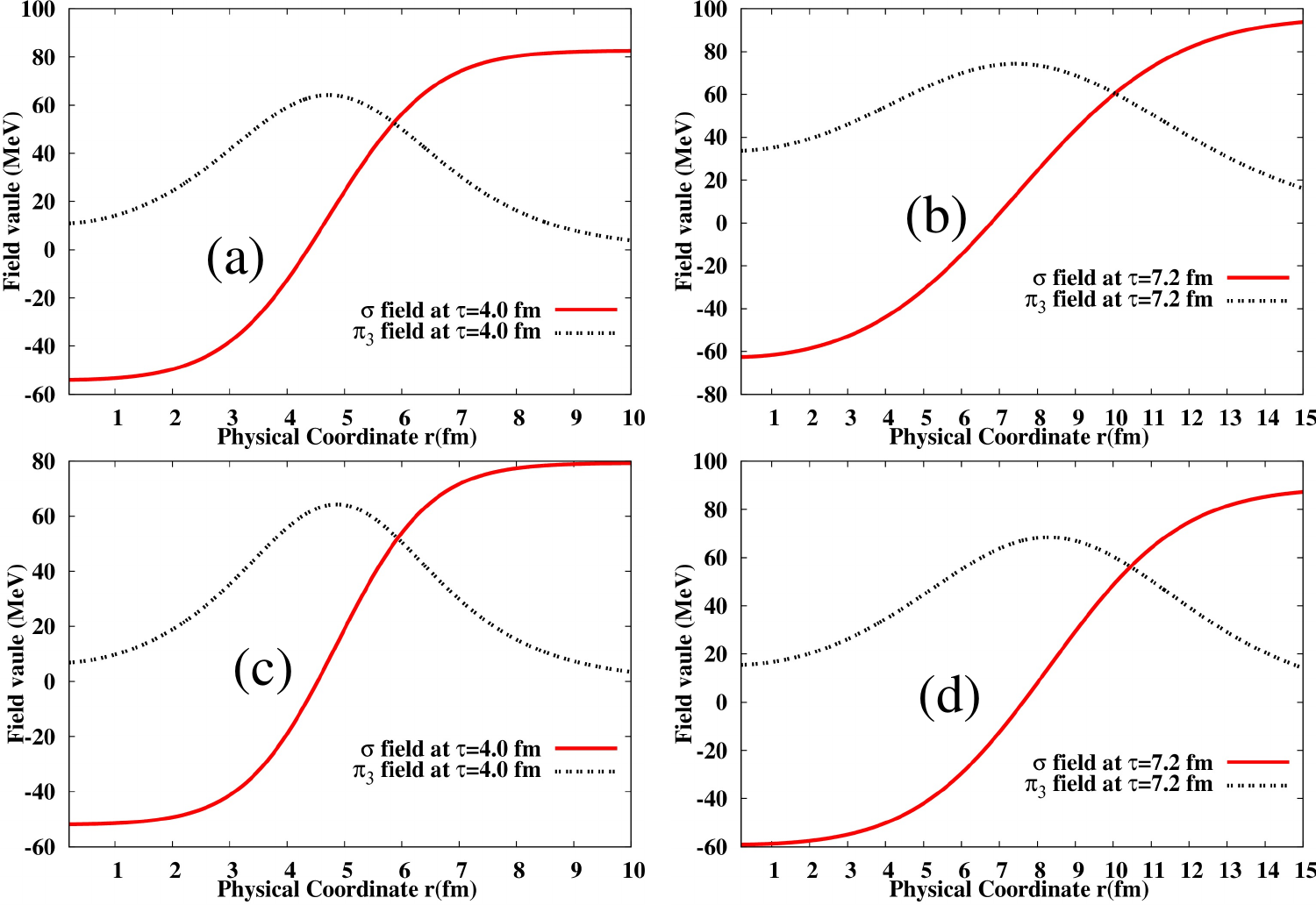}
\caption{(a) and (b) show the profiles of the chiral field at the
same stages as in Fig.2b,c and Fig.2d (starting with the same initial
profiles as in Fig.2a). $\eta^\prime$ is taken as constant for this
case with value 20 fm$^{-1}$. Comparison with Fig.2 shows that 
DCC domain stretches to a much larger size in this case.
(c) and (d) show similar stages as in (a) and (b), but now
with even larger $\eta^\prime = $ 40 fm$^{-1}$. We see much larger
DCC domain resulting here.}
\label{fig3}
\end{center}
\end{figure}

 We had shown in ref.\cite{qcdrd} that the propagating front 
solutions we obtain are very robust and almost independent of the 
initial profile of the front taken. Thus our results obtained here
are not very sensitive to the exact initial profile of the chiral 
field taken. A different profile would still lead to similar qualitative
features of the evolution of the parton system and hence a DCC domain. 

\section{CONCLUSIONS}

 We conclude by pointing out the important features of our
analysis. We focus on high multiplicity pp collisions at LHC energy
as potentially important for possible formation of {\it single} DCC 
domains. This is in contrast to heavy-ion collisions where necessarily 
one gets multiple DCC domains where clean signature of coherent pions
becomes difficult to detect. The problem of small size for pp
collision (hence small DCC domain) is circumvented by showing
the existence of slowly moving fronts governed by reaction-diffusion
equation. This delays the roll down of the disoriented chiral field
to the true vacuum significantly, while the system undergoes
rapid three dimensional expansion. This leads to stretching of
the initial DCC domain to a size of several fm which can lead
to relatively clean signals of coherent pion emission.

  Specific assumptions made in our model, such as the value of
dissipation constant, initial profile, etc. are not expected to
significantly change the main aspects of our results. In view
of our previous results in ref.\cite{qcdrd}, the existence of
slowly moving propagating front results under varied conditions and
with widely different initial profiles. This is in complete contrast
to the usual expectation that the field should rapidly roll down
to the true vacuum. Thus, high multiplicity pp collisions at LHC
energy may be an ideal place to look for the long sought signatures
of disoriented chiral condensates. The considerations presented
here are about classical evolution of the chiral field in an expanding
domain. As we mentioned above, considerations of thermal fluctuations 
and quantum decay of the DCC domain into pions will put the final 
constraint on the growth of DCC domains in our model.   

\section*{Acknowledgment}
We are very thankful to Shreyansh S. Dave for useful discussions.

\end{document}